\documentclass[aps,prd,twocolumn,preprintnumbers,superscriptaddress,dblfloatfix,nofootinbib]{revtex4-1}
\usepackage[utf8]{inputenc}
\usepackage{lmodern}
\usepackage{amsmath,amssymb,mhchem}
\usepackage{graphicx}
\usepackage{url}
\usepackage{color}
\usepackage{enumitem}
\usepackage{subfigure}
\usepackage[dvipsnames]{xcolor}
\usepackage[colorlinks=true,breaklinks=true]{hyperref}
\hypersetup{allcolors=[rgb]{0.0 0.0 0.6},linkcolor=[rgb]{0.75 0.05 0.05}}
\usepackage{bm}
\usepackage{epsfig}
\usepackage{amsmath}
\usepackage{amssymb}
\usepackage{slashed}
\usepackage{color}
\usepackage{accents}
\usepackage{url}
\usepackage[dvipsnames]{xcolor}

\newcommand{\exclude}[1]{}

\begin{document}

\preprint{CERN-TH-2020-159} 
\preprint{IPMU21-0004}

\title{Gas Heating from Spinning and Non-Spinning Evaporating Primordial Black Holes}

\author{Ranjan Laha}\email{ranjan.laha@cern.ch}
\affiliation{Theoretical Physics Department, CERN, 1211 Geneva, Switzerland}
\affiliation{Centre for High Energy Physics, Indian Institute of Science, Bangalore 560012, India}

\author{Philip Lu}\email{philiplu11@gmail.com}
\affiliation{Department of Physics and Astronomy, University of California, Los Angeles \\ Los Angeles, California, 90095-1547, USA}

\author{Volodymyr Takhistov}\email{volodymyr.takhistov@ipmu.jp}
\affiliation{Department of Physics and Astronomy, University of California, Los Angeles \\ Los Angeles, California, 90095-1547, USA}
\affiliation{Kavli Institute for the Physics and Mathematics of the Universe (WPI), UTIAS \\The University of Tokyo, Kashiwa, Chiba 277-8583, Japan}
 
\date{\today}

%%%%%%%%%%%%%%%%%%%%%%%%%%%%%%%%%%%
\begin{abstract}
Primordial black holes (PBHs) from the early Universe constitute a viable dark matter (DM) candidate and can span many orders of magnitude in mass. Light PBHs with masses around $10^{15}$ g contribute to DM and will efficiently evaporate through Hawking radiation at present time, leading to a slew of observable signatures. The emission will deposit energy and heat in the surrounding interstellar medium. We revisit the constraints from dwarf galaxy heating by evaporating non-spinning PBHs
and find that conservative constraints from Leo T dwarf galaxy are significantly weaker than previously suggested. Furthermore, we analyse gas heating from spinning evaporating PBHs. The resulting limits on PBH DM abundance are found to be stronger for evaporating spinning PBHs than for non-spinning PBHs.
\end{abstract}
%%%%%%%%%%%%%%%%%%%%%%%%%%%%%%%%%%% 
\maketitle
%%%%%%%%%%%%%%%%%%%%%%%%%%%%%%%%%%%

\section{Introduction}

Primordial black holes (PBHs), formed in the early Universe prior to any galaxies and stars, are a viable candidate for DM~(e.g.,~\cite{Zeldovich:1967,Hawking:1971ei,Carr:1974nx,Chapline:1975ojl,Meszaros:1975ef,Carr:1975qj,GarciaBellido:1996qt,Khlopov:2008qy,Frampton:2010sw,Bird:2016dcv,Kawasaki:2016pql,Carr:2016drx,Inomata:2016rbd,Pi:2017gih,Inomata:2017okj,Garcia-Bellido:2017aan,Inoue:2017csr,Georg:2017mqk,Inomata:2017bwi,Kocsis:2017yty,Ando:2017veq,Cotner:2016cvr,Cotner:2016dhw,Cotner:2017tir,Cotner:2019ykd,Cotner:2018vug,Sasaki:2018dmp,Carr:2018rid,Flores:2020drq,Deng:2017uwc,Kusenko:2020pcg}). Depending on formation, PBHs surviving until the present can span many orders of magnitude in mass, from $\sim 10^{15}$~g to well over $10^{10}\,M_{\odot}$.  They can account for the entirety of the DM in the mass window $\sim 10^{-16} - 10^{-10}\,M_{\odot}$, where there are no observational constraints~\cite{Katz:2018zrn,Smyth:2019whb,Montero-Camacho:2019jte,Dasgupta:2019cae,Carr:2020gox,Carr:2020xqk,Green:2020jor}. While significant attention has been devoted to larger mass PBHs, it has been realized recently that light PBHs can result in a larger variety of observable signatures than previously thought and is thus ripe for further exploration.

Light PBHs with mass $\lesssim 10^{-16}M_{\odot}$ existing at present time will be evaporating and copiously emitting particles through Hawking radiation~\cite{Hawking:1974sw}. Non-rotating PBHs with masses below $2.5 \times 10^{-19} M_{\odot}$ have lifetimes smaller than the age of the Universe and thus do not contribute to DM abundance~\cite{Page:1976df,Page:1977um}. Particle emission from currently evaporating PBHs produces a variety of signatures,
providing insight into this region of PBH DM parameter space. Leading constraints on light PBHs have been obtained from observations of photon flux~\cite{Carr:2009jm,Ballesteros:2019exr,Arbey:2019vqx, Laha:2020ivk}, cosmic microwave background\,\cite{Clark:2016nst,Stocker:2018avm,Poulter:2019ooo,Acharya:2020jbv}, electron and positron cosmic rays~\cite{Boudaud:2018hqb},
511 keV gamma-ray line~\cite{Okele:1980kwj,okeke1980primary,MacGibbon:1991vc,Bambi:2008kx,DeRocco:2019fjq,Laha:2019ssq,Dasgupta:2019cae}, as well as neutrinos~\cite{Dasgupta:2019cae}.  

Usually, PBHs are assumed to be non-rotating (Schwarzschild)~\cite{Chiba:2017rvs,DeLuca:2019buf,Mirbabayi:2019uph}. However, PBHs can be formed\footnote{Heavier PBHs can also efficiently acquire spin via accretion~\cite{DeLuca:2020bjf}.} with significant spin (Kerr BHs)~\cite{Harada:2016mhb,Kokubu:2018fxy,Cotner:2016cvr,Cotner:2016dhw,Cotner:2017tir,Cotner:2019ykd,Cotner:2018vug}. BH spin will affect the Hawking radiation, generally increasing the emission while favoring particles with larger spin~\cite{Hawking:1974sw,Page:1976ki,Page:1977um,Taylor:1998dk}.
Furthermore, the mass limit of $\sim$ $2.5 \times 10^{-19} M_{\odot}$ for PBHs below which their lifetime is smaller than the age of the Universe varies by a factor of $\sim 2$ for maximally rotating PBHs \cite{Page:1976ki,Dong:2015yjs,Arbey:2019jmj}.
Besides mass (and electric charge), angular momentum constitutes a fundamental conserved parameter of a BH. Hence, it is important to explore the implications of spin for observations~\cite{Arbey:2019vqx,Dong:2015yjs,Kuhnel:2019zbc,Arbey:2019jmj,Bai:2019zcd,Dasgupta:2019cae}.  

Recently, observations from dwarf galaxies (in particular, Leo T) have been used to constrain stellar and intermediate-mass
PBHs by considering heating of interstellar medium (ISM) gas due to PBH interactions~\cite{Lu:2020bmd}. This represents a new signature not previously considered for PBHs. Subsequently, Ref.~\cite{Kim:2020ngi}
considered heating of ISM gas due to light evaporating non-rotating PBHs.
 
In this work, we revisit and provide an alternative treatment of gas heating due to evaporating PBHs, focusing on the dwarf galaxy Leo T. We find that a more detailed, conservative, and
proper treatment of energy deposition from PBH emission results in significantly weaker constraints than reported in the analysis of Ref.\,\cite{Kim:2020ngi}. Furthermore, we study gas heating due to evaporating PBHs with significant spin.

\begin{figure*}[tb]
\begin{center}
\includegraphics[trim={0mm 0mm 0 0},clip,width=.32\textwidth]{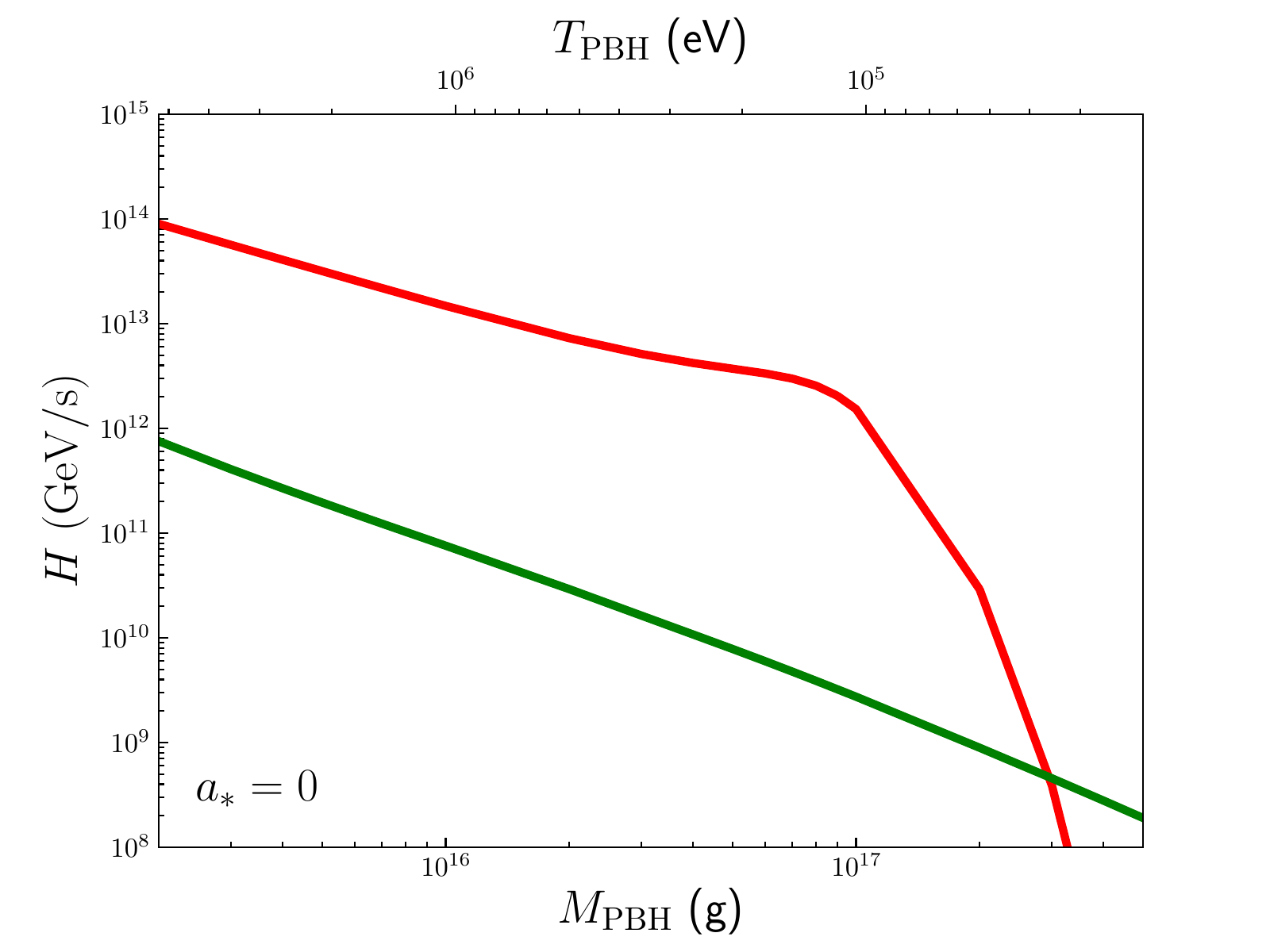}
\includegraphics[trim={0mm 0mm 0 0mm},clip,width=.32\textwidth]{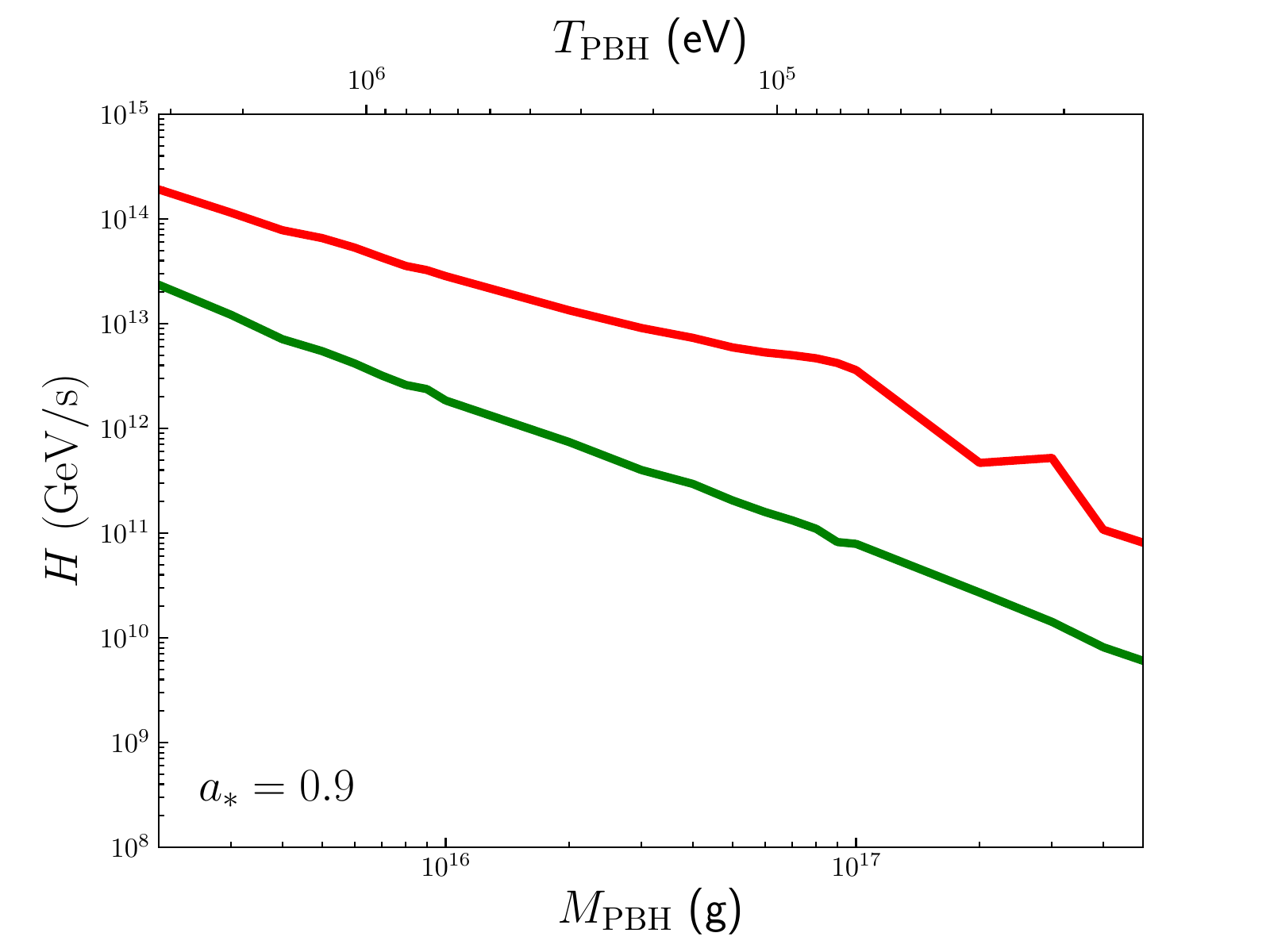}
\includegraphics[trim={0mm 0mm 0 0mm},clip,width=.32\textwidth]{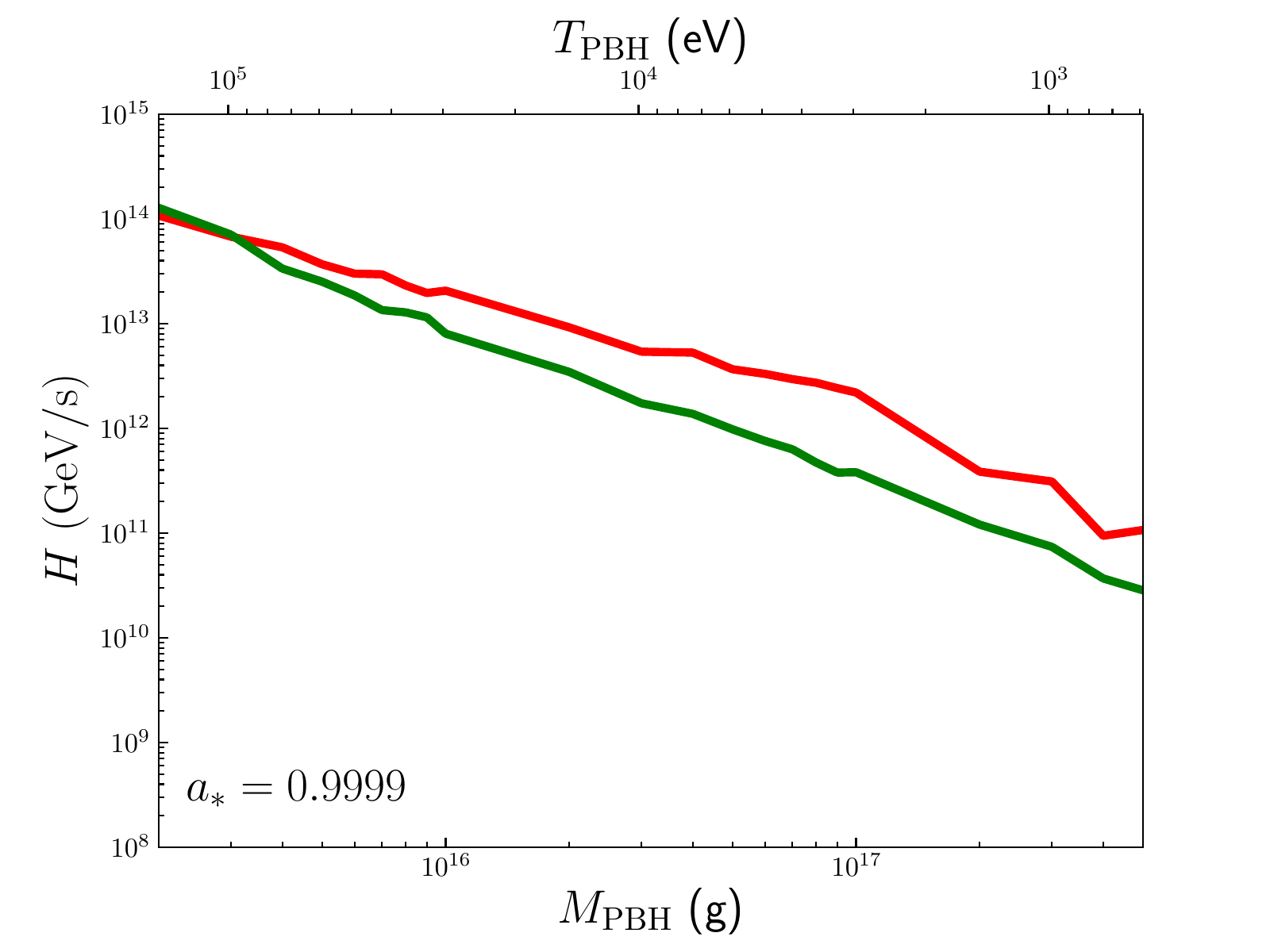}
\caption{ \label{fig:heatpbh} 
Emission components from evaporating PBHs contributing to gas heating in Leo T, assuming PBHs are non-rotating with $a_* = 0$ \textbf{[Left]}, PBHs are spinning with $a_* = 0.9$ \textbf{[Middle]}, and PBHs approaching the Kerr BH limit with spin $a_* = 0.9999$ \textbf{[Right]}. Contributions from primary photons (green line) and electrons/positrons (red line) are displayed.
}
\end{center}
\end{figure*}
 
This study is organized as follows. In Sec.~\ref{sec:pbhevap}, we discuss emission and generate the spectrum for spinning and non-spinning PBHs. In Sec.\,\ref{sec:leot}, we describe properties of target system Leo T dwarf galaxy. In Sec.~\ref{sec:gasheat}, we discuss the energy deposition and heating in dwarf galaxies from evaporating PBHs, focusing on Leo T. In Sec.~\ref{sec:gascool}, we discuss gas cooling and thermal balance with heating. Finally, we summarize in Sec.~\ref{sec:conclusions}.

\vspace{-3mm}

\section{Evaporating Black Hole Emission}
\label{sec:pbhevap}

An un-charged\footnote{BHs with mass below $\lesssim 10^5 M_{\odot}$ 
are expected to rapidly lose any accumulated charge due to Schwinger pair production~\cite{Gibbons:1975kk,Zaumen:1975je}.} rotating (Kerr) PBH radiates at a temperature given by~\cite{Page:1976df,Page:1976ki,Page:1977um,MacGibbon:1990,MacGibbon:1991,MacGibbon:2007yq}
\begin{equation}
\label{eq:temperature}
    T_{\rm PBH} = \dfrac{1}{4 \pi G M_{\rm PBH}} \Big(\dfrac{\sqrt{1-a_*^2}}{1 + \sqrt{1-a_*^2}}\Big)~,
\end{equation}
where $G$ is the gravitational constant, $M_{\rm PBH}$ and $a_* = J_{\rm PBH}/(G M_{\rm PBH}^2)$ are the PBH mass and reduced spin Kerr parameter, for a PBH with angular momentum $J_{\rm PBH}$.
In the limit $a_*\rightarrow 0$, Eq.~\eqref{eq:temperature} reduces to the usual Hawking evaporation temperature of a Schwarzschild BH,
$T_{\rm PBH} \simeq 1.1\textrm{ MeV}\left(M_{\rm PBH}/10^{16} \text{g}\right)^{-1}$. 
The temperature is seen to be significantly 
diminished for a Kerr BH in the limit $a_* \rightarrow 1$.

Evaporating PBHs start to emit significant quantities of a given particle as the BH temperature reaches the particle mass, and at high temperatures the emission spectrum resembles that of blackbody radiation\,\cite{Page:1976df}.
For spin-1/2 particles, the emission peak occurs at $E \simeq \, 4.03 \, T_{\rm PBH}$\,\cite{MacGibbon:2007yq}. At lower BH masses, secondary emission channels due to quark and gluon QCD jets become relevant.

For primary emission, the number of particles, $N_i$, emitted per unit energy per unit time is given by \cite{Page:1976df,Page:1976ki,Page:1977um,MacGibbon:1990,MacGibbon:1991,MacGibbon:2007yq}
\begin{equation}
\label{eq:emissionrate}
    \dfrac{d^2 N_i}{dt dE} = \dfrac{1}{2 \pi}\sum_{\rm dof}\dfrac{\Gamma_i(E,M_{\rm PBH}, a_*)}{e^{E^{\prime}/T_{\rm PBH}}\pm 1}~,
\end{equation}
where the
greybody factor $\Gamma_i(E,M_{\rm PBH}, a_*)$ encodes the probability that the emitted particle overcomes the gravitational well
of the BH, $E^{\prime}$ is the total energy of a particle when taking BH rotation into account, the $\pm$ signs are for fermions and bosons, respectively, and summation is over considered degrees of freedom. Secondary emission of particles from QCD jets can be computed numerically~\cite{MacGibbon:1990zk}.

For our study we 
generate the PBH emission spectrum for each
particle species using {\tt BlackHawk} code\footnote{Results of numerical computation have been verified against semi-analytical formulas \cite{Page:1976df,Page:1976ki,Page:1977um,MacGibbon:1990}.}~\cite{Arbey:2019mbc}.

\section{Target System: Leo T}
\label{sec:leot}

DM-rich dwarf galaxies represent favorable environments to investigate the effects of PBH heating due to interactions with gas. Throughout this work, we focus on the well-modelled Leo T dwarf galaxy as our target system due to its desirable cooling, gas, and DM properties. We stress, however, that our analysis is general and can be readily extended to other gas systems of interest\footnote{We estimate that heating of Milky Way gas clouds leads to weaker bounds.}.

To describe Leo T, we follow the model of Refs.\,\cite{2007ApJ...670..313S,Faerman:2013pmm,RyanWeber:2007fb} for the DM density, neutral hydrogen (HI) gas distribution, and ionization fraction. The hydrogen gas in the inner $r_s <350$ pc of Leo T system is largely un-ionized~\cite{Wadekar:2019xnf,Kim:2020ngi}. We consider only this central region, employing the average HI gas density of $n_{\rm H}=0.07\textrm{ cm}^{-3}$,
ionization fraction $x_e = 0.01$ and DM density of $1.75 \textrm{ GeV}\textrm{ cm}^{-3}$~\cite{2007ApJ...670..313S}.
Hence, the gas column density can be estimated as   $n_{\rm H} r_s = 9.72\times10^{19} \textrm{ cm}^{-2}$ and the mass column density as $m_{\rm H} n_{\rm H} r_s = 1.627\times10^{-4}\textrm{ g}\textrm{ cm}^{-2}$. The velocity dispersion of the HI gas in this region $\sigma_v=6.9$ km/s~\cite{RyanWeber:2007fb,Simon:2007dq,2018A&A...612A..26A} suggests a gas temperature of $T\simeq 6000$~K. In contrast to the case of PBH accretion emission analysis \cite{Lu:2020bmd}, the velocity dispersion and distribution of the HI gas and DM are not as relevant here as long as they remain non-relativistic.

\begin{figure*}[tb]
\begin{center}
\includegraphics[trim={0mm 0mm 0 0},clip,width=.49\textwidth]{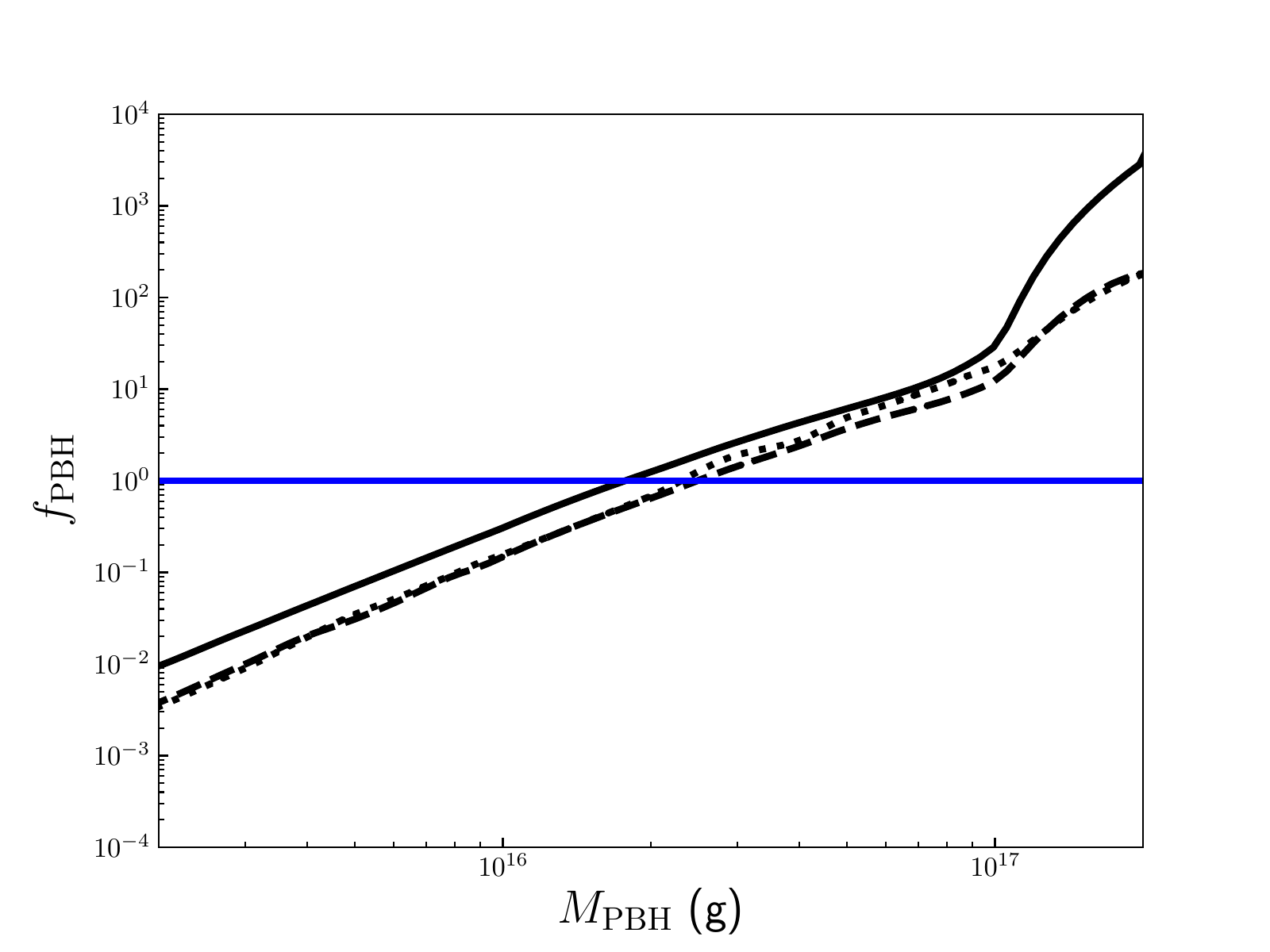}
\includegraphics[trim={0mm 0mm 0 0mm},clip,width=.49\textwidth]{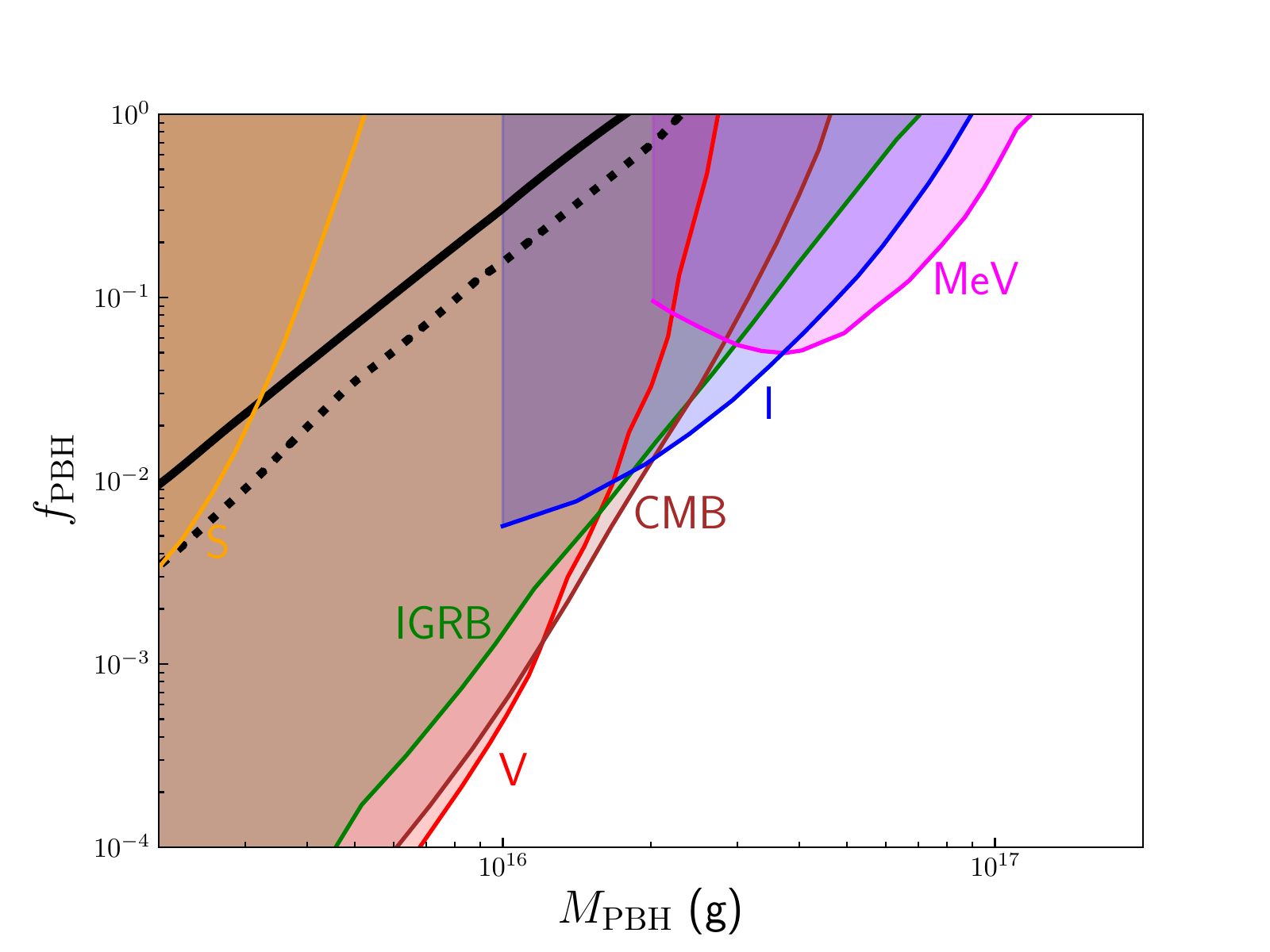}
\caption{ \label{fig:heatpbh2} 
Constraints from Leo T on the fraction of DM PBHs, $f_{\rm PBH}$, for a monochromatic PBH mass function. \textbf{[Left]} Results for non-rotating PBHs with spin $a_* = 0$ (black solid line), PBHs with spin $a_* = 0.9$ (black dashed line) and PBHs approaching Kerr limit with $a_* = 0.9999$ (black dotted line) are shown. For non-rotating PBHs, the ``elbow'' feature seen at higher PBH masses is due to exponential decrease of the electron contributions. \textbf{[Right]} Overlay of our results with existing constraints on non-spinning PBHs from \textit{Voyager-1} detection of positrons and using propagation model B without background (‘‘V", shaded red) \cite{Boudaud:2018hqb}, \textit{Planck} cosmic microwave background (‘‘CMB", shaded brown) \cite{Clark:2016nst}, isotropic gamma-ray background (‘‘IGRB", shaded green) \cite{Carr:2009jm, Ballesteros:2019exr, Carr:2020gox}, \textit{INTEGRAL} 511 keV emission line for the isothermal DM profile with 1.5 kpc positron annihilation region (‘‘I", shaded blue)~\cite{DeRocco:2019fjq,Laha:2019ssq,Dasgupta:2019cae}, \textit{Super-Kamiokande} neutrinos (‘‘S", shaded orange)~\cite{Dasgupta:2019cae}, as well as \textit{INTEGRAL} Galactic Center MeV flux (‘‘MeV", shaded magenta) \cite{Laha:2020ivk}.  The constraint marked ``I" and ``MeV" are shown till the lowest PBH masses as displayed in Refs.\,\cite{Laha:2019ssq} and \cite{Laha:2020ivk} respectively.}
\end{center}
\end{figure*}

\section{Gas Heating by Evaporating PBHs}
\label{sec:gasheat}

As proposed in Ref.\,\cite{Lu:2020bmd}, accretion emission from heavier PBHs will deposit energy and heat the gas in surrounding interstellar medium. 
It was subsequently suggested that emitted particles from evaporating light PBHs can also deposit energy and heat surrounding gas \cite{Kim:2020ngi}. Below, we revisit gas heating due to non-rotating evaporating PBHs with an improved treatment.
We also extend our study of gas heating to emission from rotating evaporating PBHs.

For our PBH masses of interest, both photon as well as electron/positron emission channels from evaporating PBHs can contribute to heating. In Ref.~\cite{Kim:2020ngi}, photon heating contribution has been assumed negligible due to power law scaling of photo-electric cross-section $\sigma_{\rm PE} \propto E^{-3.5}$ when photon energies are above keV. However, for photon energies around MeV that are typical to our study, the cumulative photon interaction cross-section levels out, primarily due to Compton scattering contribution (see Fig.\,33.19 of Ref.\,\cite{Tanabashi:2018oca}).
The average heating rate due to photon emission of PBH of mass $M_{\rm PBH}$ and spin $a_*$ is given by \cite{Lu:2020bmd}
\begin{equation}
\label{eq:photonheating}
H_\gamma (M_{\rm PBH},a_*) = \int_{0}^{\infty} f_{h}(E) E \frac{d^2 N_\gamma}{dtdE} \left(1-e^{-\tau}\right) dE~,
\end{equation}
where $f_h(E) \sim \mathcal{O}(1)$ is the fraction of photon energy loss deposited as heat, and  $\tau = m_\textrm{H} n_\textrm{H} r_s/\lambda$ is the optical depth of gas in terms of the absorption length $\lambda$. We take the cumulative photon absorption length from Ref. \cite{Tanabashi:2018oca}.
We assume that the photon  deposits heat similarly to electrons of the same energy. Hence, we approximate the fraction of energy deposited as heat to be similar to that of electrons, $f_{h}(E)=0.367+0.395(11\textrm{ eV}/(E - m_e))^{0.7}$ \cite{1985ApJ...298..268S,Ricotti:2001zf,Furlanetto:2009uf,Kim:2020ngi}, where $m_e$ is the electron mass. The efficiency of photon heating is rather poor, with the heat deposited within Leo T from characteristic MeV photons with $\lambda \simeq 10$ g/cm$^2$ being only $\sim 10^{-5}$ fraction of the photon energy. 

Analogously to the photon case, heating due to PBH electron/positron emission can be stated as
\begin{equation}
\label{eq:electronheating}
H_e (M_{\rm PBH},a_*) = 2 \int_{m_e}^{\infty} f_{h}(E) [E-m_e] \frac{d^2 N_e}{dtdE} \left(1-e^{-\tau}\right) dE~,
\end{equation}
where factor of 2 comes from summing contributions of electrons and positrons, $f_h(E)$ is taken as before and the factor $(1 - e^{-\tau})$ accounts for the gas system's optical thickness. When the system is not optically thick, optical depth can be written in terms of stopping power, $S(E)$, as $\tau \simeq m_\textrm{H} n_\textrm{H} r_s S(E)/E$. For the electron stopping power on hydrogen gas, we use NIST database \cite{NISTdatabase}. For characteristic MeV electrons with $S(E)\simeq 2$ MeV cm$^2$/g, only $\sim 10^{-4}$ fraction of the electron energy is deposited as heat in Leo T\footnote{We note that inclusion of a small hydrogen ionization fraction does not significantly affect these considerations~\cite{radjour2016}.}. 

In Ref. \cite{Kim:2020ngi} it was argued that for Leo T MeV scale energies are efficiently thermalized by elastic scattering within
the cooling time scale.
However, the suppression of gas heating in Eq.\,\eqref{eq:electronheating} from positrons and electrons, $(1 - e^{-\tau})$, due to optical thickness of the gas system was not accounted for in the study of Ref. \cite{Kim:2020ngi}, effectively assuming that the gas is fully optically thick (i.e., $\tau \gtrsim 1$). This lead to significant overestimation of the limits from evaporating PBH gas heating than we find, as discussed below. Since the size of the gas system in consideration is finite and it can be optically thin, this factor should be present. We note that presence of magnetic fields in the ISM can also affect emitted positrons. However, the strength, orientation and distribution of magnetic fields in Leo T is highly uncertain and very poorly known. Further, propagation of positrons can be affected in a non-trivial way by diffusion, collisions, advection, and other processes. Even in a relatively well-studied region like the Milky Way Galactic Center, the positron propagation distance is highly uncertain~\cite{Panther:2018xvc}.  We expect such uncertainties to be present in Leo T too, especially in the absence of our understanding of the magnetic field, turbulence, and other astrophysical properties of that galaxy.  Hence, our resulting bounds from PBH heating are conservative.

In Fig.~\ref{fig:heatpbh} (left panel), we display the resulting heating rates $H (M_{\rm PBH}, a_*)$ for Leo T, including contributions of primary photons and electrons/positrons. Secondary emission of photons and electrons/positrons is negligible in our range of interest $M_{\rm PBH}\gtrsim 2\times10^{15}$ g, and hence is not shown\footnote{As discussed in recent work of Ref.~\cite{Coogan:2020tuf}, computation of secondary production by {\tt BlackHawk} can be improved over some emission regimes.}. Electrons/positrons are seen to provide the dominant contribution to heating rate within a broad range of parameter space of interest. Photons provide a sub-dominant contribution to the heating, but could in fact dominate in the regimes $T_{\rm PBH} \ll m_e$ (where electron emission is heavily suppressed). 

We further analyze heating from spinning PBHs, displaying results for $a_* = 0.9$ (Fig.~\ref{fig:heatpbh}, middle panel) and $a_* = 0.9999$ (Fig.~\ref{fig:heatpbh}, right panel). The emission for $a_* = 0$ and $a_* = 0.9$ is seen to be similar. As the spin approaches the extreme Kerr limit, $a_* \rightarrow 1$, the pattern of PBH emission and hence heating contributions changes. The emission tends to be higher for spinning PBHs and for highly spinning PBHs, photons can become dominant at smaller PBH masses, as they are produced in greater abundance than electrons~\cite{Page:1976ki}.

\section{Cooling and Thermal Balance}
\label{sec:gascool} 

The thermal balance of heating from PBHs and gas cooling allows us to constrain the PBH abundance with Leo T\,\cite{Kim:2020ngi,Lu:2020bmd}. Use of gas heating to constrain particle DM has been suggested in Ref.~\cite{Bhoonah:2018gjb,Bhoonah:2018wmw,Wadekar:2019xnf}. We ignore the possible additional contributions of natural heating sources, resulting in more conservative bounds.

Gas temperature exchange is a complex process and a detailed analysis involving a full chemistry network can be performed using numerical  methods~\cite{Smith:2016hsc}. For the  parameters of interest, we employ the approximate gas cooling rate results obtained in Ref.~\cite{Wadekar:2019xnf}. The cooling rate per unit volume of the hydrogen gas is given by
\begin{equation}
\label{eq:coolinggen}
    \dot{C} = n_H^2 10^{[\textrm{Fe/H}]}\Lambda(T)\,,
\end{equation}
where [Fe/H]$ \equiv \log_{10}(n_{\rm Fe}/n_{\rm H})_{\rm gas} - \log_{10}(n_{\rm Fe}/n_{\rm H})_{\rm Sun}$ is the metallicity, and 
$\Lambda(T)$ is the cooling function.  We obtain $\Lambda(T) = 2.51\times10^{-28}T^{0.6}$, valid for $300~\text{K} < T < 8000~\text{K}$~\cite{Wadekar:2019xnf}, via a numerical fit to the results of Ref.~\cite{Smith:2016hsc}. Following the analysis of Refs.~\cite{Wadekar:2019xnf,Lu:2020bmd}, we adopt $\dot{C}=2.28\times10^{-30}\textrm{ erg}\textrm{ cm}^{-3}\textrm{ s}^{-1}$ for the cooling rate in Leo T.

From the PBH DM fraction, $f_\textrm{PBH}$, and average DM density in Leo T, $\rho_\textrm{DM}\simeq 1.75\textrm{ GeV}\textrm{ cm}^{-3}$, the total number of PBHs residing in Leo T is 
\begin{equation}
  N_\textrm{PBH}=\Big(\dfrac{4\pi r_s^3}{3}\Big)\dfrac{f_\textrm{PBH}\rho_\textrm{DM}}{M_{\rm PBH}}.
\end{equation}
We take the average density such that the $N_{\rm PBH}$ is the same as that obtained while integrating over the DM profile.
Requiring the total generated heat,
$N_{\rm PBH} H(M_{\rm PBH},a_*)$, to be less than the total cooling rate in the central region of Leo T, yields the constraint~\cite{Lu:2020bmd}
\begin{equation}
\label{eq:constraint}
    f_\textrm{PBH}<f_\textrm{bound} = \frac{M_{\rm PBH}\dot{C}}{\rho_\textrm{DM}H(M_{\rm PBH},a_*)}~.
\end{equation}

In Fig.\,\ref{fig:heatpbh2}, we display our resulting constraints from PBH gas heating along with other existing limits, assuming a monochromatic PBH mass-function.  Our results can be readily extended for other PBH mass-functions.  Spinning PBHs are seen to induce stronger limits than non-spinning PBHs. Our results are several orders of magnitude below the results suggested by the analysis of Ref. \cite{Kim:2020ngi}, which can be attributed primarily 
to not accounting for the optical thickness of gas as described above. Furthermore, we have extended the constraints to smaller PBHs masses.  

A better understanding of the standard astrophysical heating rate in Leo T can substantially improve this limit.  Similarly, discovery of more DM dominated dwarf galaxy systems and a good understanding of heating and cooling rates inside them can even lead to discovery of low-mass PBHs via this technique.

\section{Conclusions}
\label{sec:conclusions}
  
Light PBHs, with masses $\lesssim 10^{17}$ g, contributing to DM will significantly emit particles via Hawking radiation depositing energy and heat in the surrounding gas. We have studied gas heating due to spinning and non-spinning PBHs, focusing on the dwarf galaxy Leo T.
A detailed, conservative, and proper treatment of heating results in presented limits being significantly weaker than previously claimed. We find that limits from spinning evaporating PBHs are stronger than for the non-spinning case.

\section*{Acknowledgments}
We thank Jeremy Auffinger, Hyungjin Kim, Alexander Kusenko, and Anupam Ray for comments and discussions.  The work of P.L. and V.T. was supported by the U.S. Department of Energy (DOE) Grant No. DE-SC0009937. R.L. thanks CERN theory group for support. V.T. was also supported by the World Premier International Research Center Initiative (WPI), MEXT, Japan.

\bibliographystyle{bibi}
\bibliography{bibliography}
 
\end{document}